\documentclass{elsart}

\usepackage{natbib}

\usepackage{graphicx}

\usepackage{amssymb}

\begin{document}

\begin{frontmatter}



\title{Chaotic maps coupled with random delays: connectivity, topology,
and network propensity for synchronization}


\author[lab1]{Arturo C. Mart\'{\i}},
\ead{marti@fisica.edu.uy}
\author[lab1]{Marcelo Ponce C.} and
\ead{mponce@fisica.edu.uy}
\author[lab1,lab2]{Cristina Masoller}
\ead{cris@fisica.edu.uy}

\address[lab1]{Instituto de F\'{\i}sica, Facultad de
  Ciencias, Universidad de la Rep\'ublica, Igu\'a 4225, Montevideo
  11400, Uruguay}

\address[lab2]{Departament de Fisica i Enginyeria
  Nuclear, Universitat Politecnica de Catalunya, Colom 11, E-08222
  Terrassa, Spain}

\begin{abstract}
We study the influence of network topology and connectivity on the
synchronization properties of chaotic logistic maps, interacting with
random delay times. Four different types of topologies are investigated:
two regular (a ring-type and a ring-type with a central node) and two
random (free-scale Barabasi-Albert and small-world Newman-Watts). The
influence of the network connectivity is studied by varying the average
number of links per node, while keeping constant the total input that
each map receives from its neighbors. For weak coupling the array does
not synchronize regardless the topology or connectivity of the network;
however, for certain connectivity values there is enhanced
coherence. For strong coupling the array synchronizes in the homogeneous
steady-state, where the chaotic dynamics of the individual maps is
suppressed. For both, weak and strong coupling, the array propensity for
synchronization is largely independent of the network topology and
depends mainly on the average number of links per node.
\end{abstract}

\end{frontmatter}

The emergence of dynamical order in complex systems has
been widely studied during the last years \cite{review,damian,boccaletti06}.
One important issue is the propensity for synchronization of networks of
dynamical elements. In this context, coupled maps \cite{kaneko} are
excellent tools for understanding the mechanisms of emergency of
synchrony and collective behavior in complex systems composed of
mutually coupled nonlinear units. Not only from an academical point of
view but also from an applied perspective, cooperative behavior arises
in many fields of science and classical examples include the onset of
rhythmic activity in the brain, the flashing on and off in unison of
populations of fireflies, the emission of chirps by populations of
crickets and the synchronization of laser arrays and Josephson junctions
\cite{review}. Coupled map lattices have proven to be a useful tool
because by simplifying the dynamics of the individual units it is
possible to simulate  large ensembles of coupled units.

Network synchronizability and its relation with the topology has
recently received a great deal of attention. F. Atay et
al. \cite{atay_PRL_2004} found, in the case of fixed (constants) delays
that scale-free and random networks exhibit better synchronization
properties than regular networks. More recently, A. Motter et
al. \cite{motter2005} identified that the synchronization of complex
networks follows a diffusive mechanism where the mean connectivity plays
a key role.  In a previous paper \cite{prl} we investigated the relation
of the topology with the ability to synchronize under the presence of
random delays.  We found that in this case the synchronization
properties depend largely on the mean connectivity of the
network. However, the topology does not play an important role.  The aim
of this paper is to further investigate this point.

The evolution equations for $N$ coupled logistic map with random delays
are
\begin{equation}
\label{mapa} x_i(t+1)= (1-\epsilon) f[x_i(t)] + \frac{\epsilon}
{b_i} \sum_{j=1}^N \eta_{ij} f[x_j(t-\tau_{ij})].
\end{equation}
Here $t$ is a discrete time index, $i$ is a discrete spatial index
($i=1\dots N$), $f(x)=ax(1-x)$ is the logistic map, the matrix
$\eta=(\eta_{ij})$ defines the connectivity of the array:
$\eta_{ij}=\eta_{ji}=1$ if there is a link between the $i$th and
$j$th nodes, and zero otherwise. $\epsilon$ is the coupling strength
and $\tau_{ij}$ is the delay time in the interaction between the
$i$th and $j$th nodes (the delay times $\tau_{ij}$ and $\tau_{ji}$
need not be equal). The sum in Eq.~(\ref{mapa}) runs over the $b_i$
nodes which are coupled to the $i$th node ($b_i = \sum_j
\eta_{ij}$). The normalized pre-factor $1/b_i$ means that each map
receives the same total input from its neighbors.

In a previous work \cite{prl} we found that if the delays $\tau_{ij}$
are random enough, for adequate coupling strength the array synchronizes
in the {\it spatially homogeneous steady-state}, $x_i(t)=x_0$ for all $i$,
where $x_0$ is the nontrivial fixed point, $x_0=1-1/a$ \cite{prl}.  This
synchronization behavior is in contrast with the behavior with fixed
delays [if $\tau_{ij}=\tau_0$ $\forall$ $i$, $j$, the array synchronizes
in a spatially homogeneous {\it time-dependent state}, where the
dynamics is either periodic or chaotic depending on $\tau_0$
\cite{atay_PRL_2004}], and can be understood in terms of the analogy
between globally coupled maps and a single map with a external driving
\cite{cosenza2,marti}.

To investigate the influence of the topology we consider four
networks, two of them are regular and the other two are random. The
regular ones are a ring of nearest-neighbor elements while in the second
one we added a central node connected to all other nodes. The random
networks consist of a scale free network constructed according to the
Barabasi-Albert method and, concerning the last one, we use the
small-world topology proposed by Newman and Watts.  To characterize the
transition to synchronization we use the indicator $\sigma^2 = 1/N
\langle \sum_i [x_i(t)- \langle x \rangle]^2 \rangle_t$, where
$\langle.\rangle$ denotes an average over the elements of the array and
$\langle . \rangle_t$ denotes an average over time. If the array
synchronizes in a spatially homogeneous state, $x_i(t)=x_j(t)$ $\forall$
$i$, $j$, and, obviously, $\sigma^2 =0$.

\begin{figure}
\begin{center}
\includegraphics[width=12.5cm]{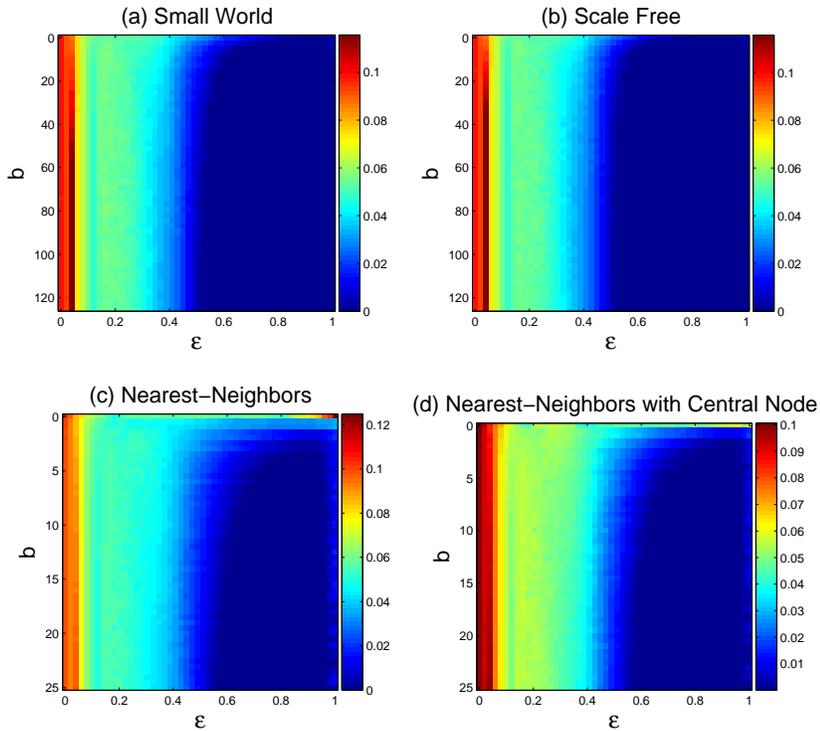}
 \end{center}
\caption{(Color online) Synchronization regions for the four different
networks considered. The density plots represent the parameter
 $\sigma^2$ as a function of $\epsilon$ and $b$ ($N=500$ and $a=4.$). }
\label{fig-cpr}
\end{figure}

\begin{figure}
\begin{center}
\includegraphics[width=4.5cm,height=6.1cm]{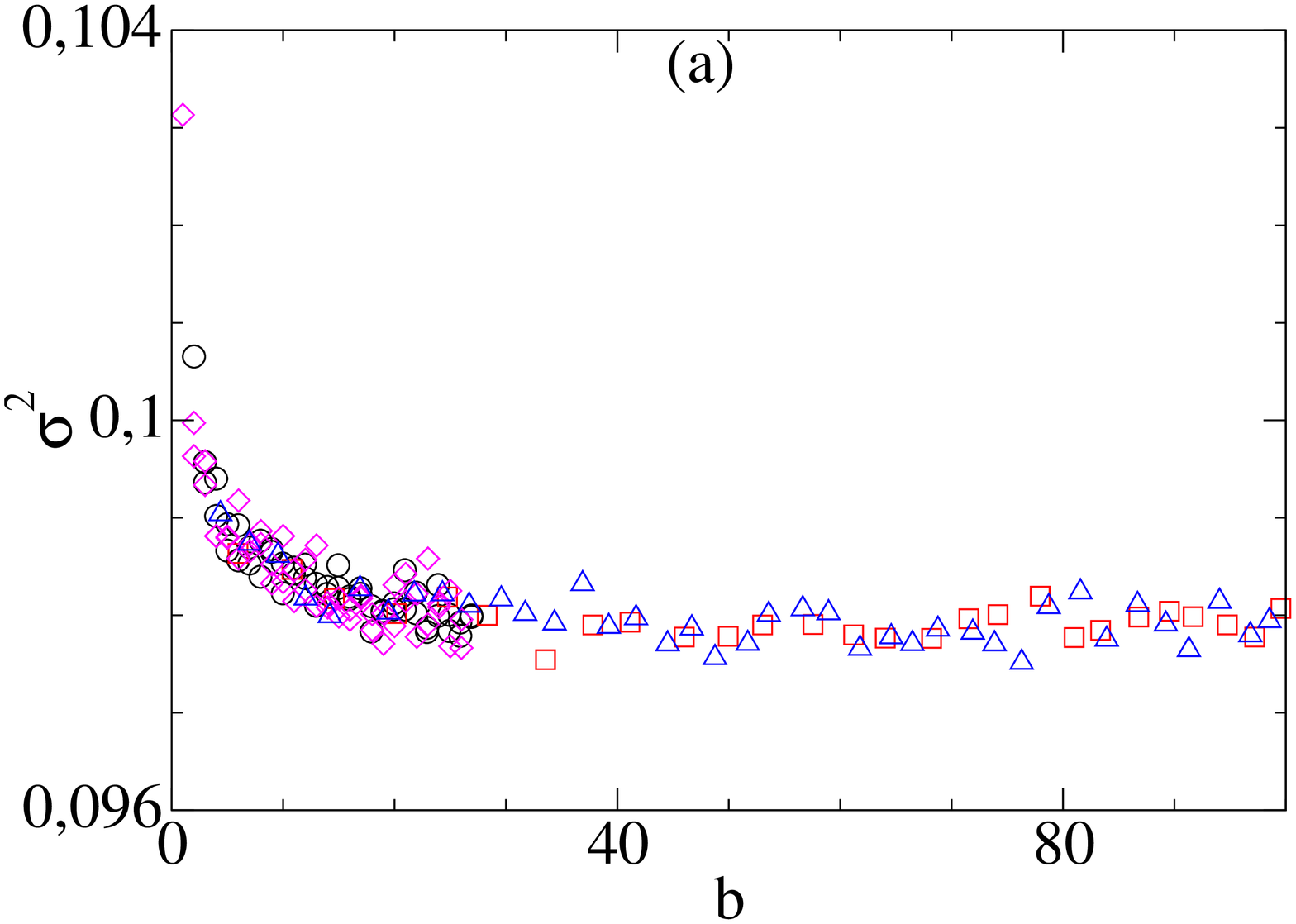}
\includegraphics[width=4.5cm,height=5.94cm]{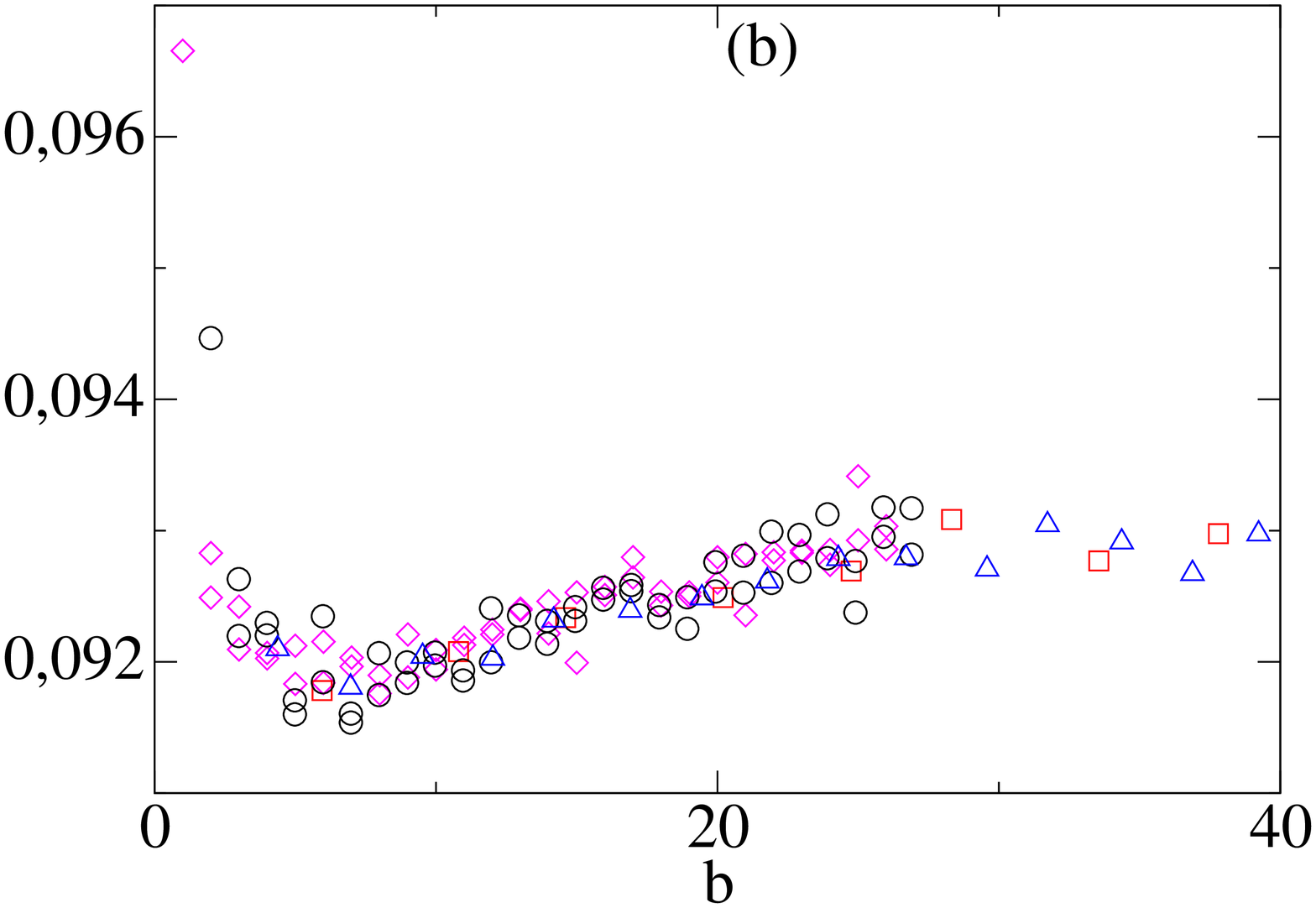}
\includegraphics[width=4.5cm,height=6.1cm]{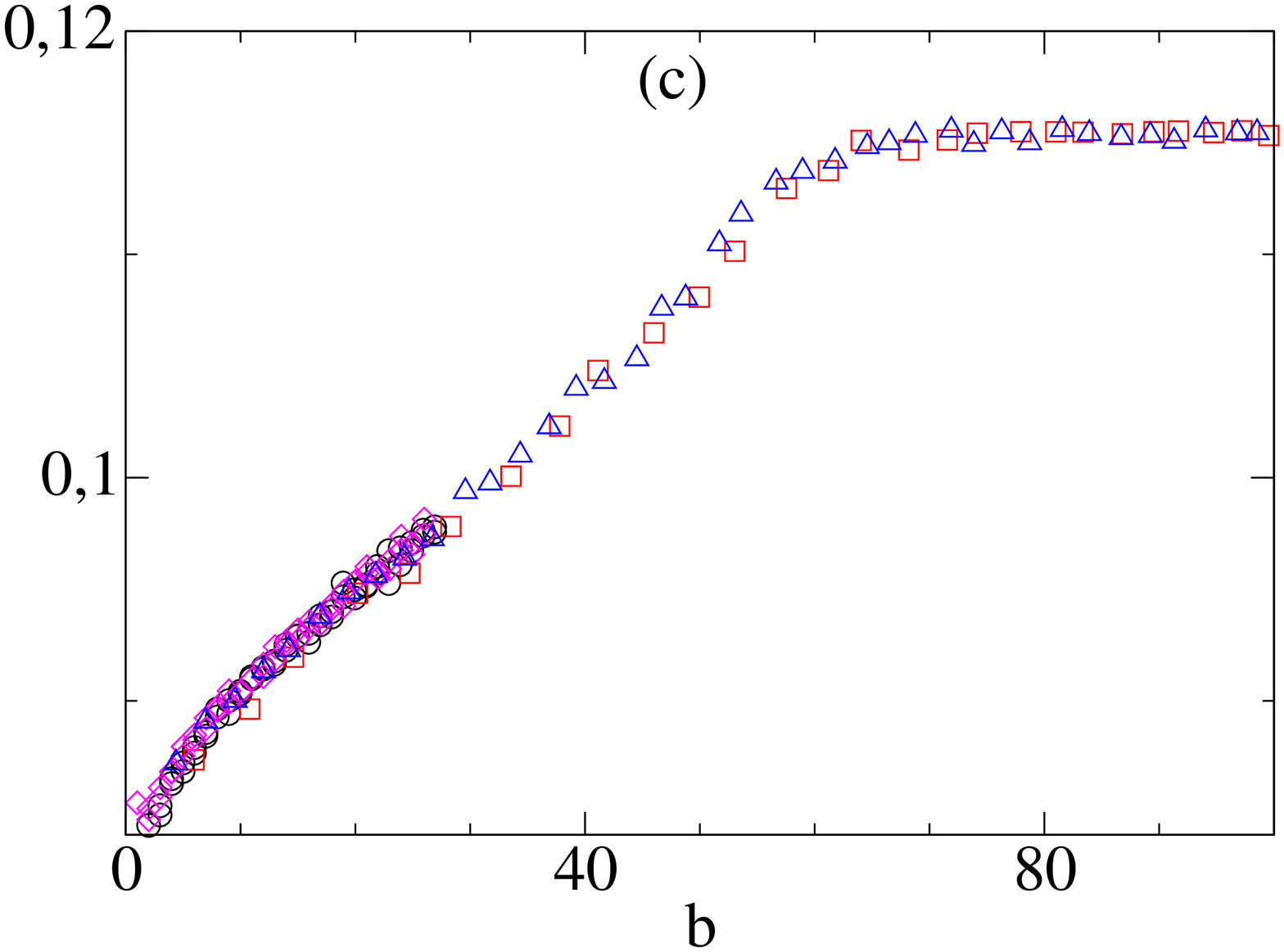}
\end{center}
\caption{ $\sigma^2$ as a function of the number of neighbors $b$ for
the weak coupling regime and different topologies: nearest-neighbors
with central node ($\circ$), scale-free ($\Box$), nearest-neighbors
($\Diamond$) and small-world ($\triangle$).  Parameters are: $N=500$,
$a=4$,  (a) $\epsilon=0.02$, (b) $\epsilon=0.04$ and (c)
$\epsilon=0.06$ } \label{fig-weak}
\end{figure}

\begin{figure}
\begin{center}
\includegraphics[width=4.5cm,height=5.5cm]{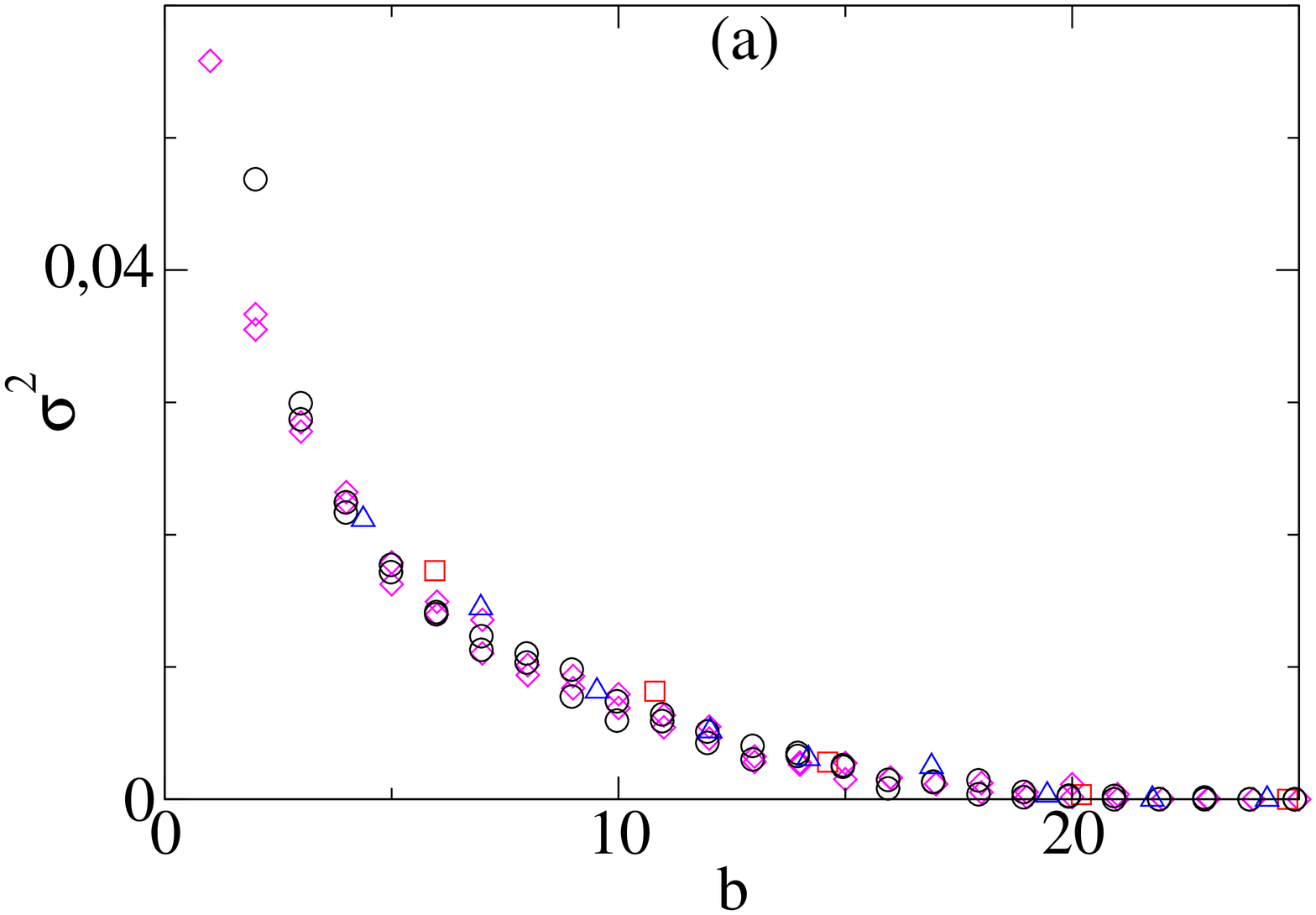} 
\includegraphics[width=4.5cm,height=5.5cm]{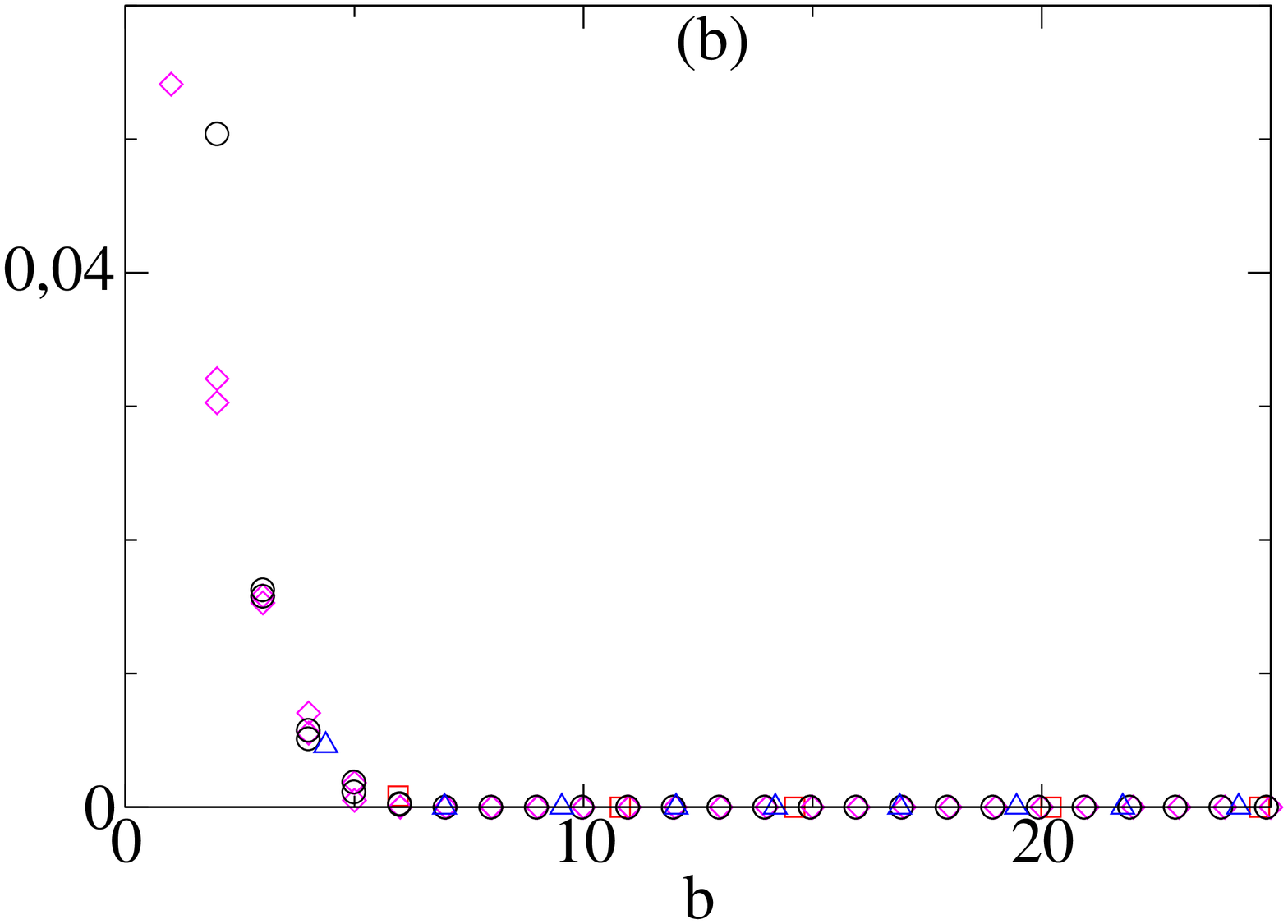} 
\includegraphics[width=4.5cm,height=5.5cm]{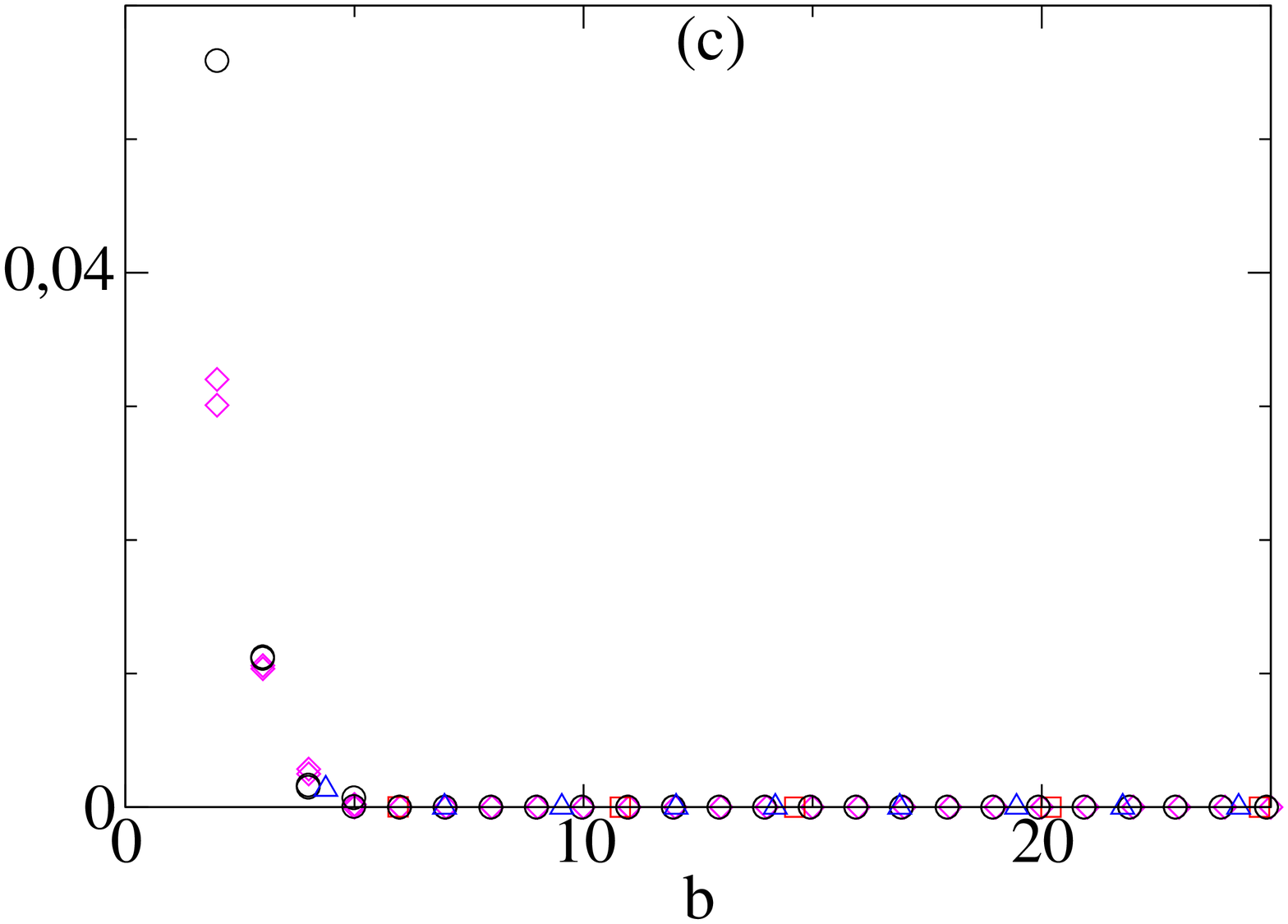} 
\end{center}
\caption{ $\sigma^2$ as a function of the number of neighbors for the
 strong coupling regime and different topologies.  (a) $\epsilon=0.6$,
 (b) $\epsilon=0.8$ and (c) $\epsilon=0.9$.  Same symbols and
 parameters values as Fig.~\ref{fig-weak}.}  \label{fig-strong}
\end{figure}

We consider Gaussian distributed delays: $\tau_{ij} = \tau_0 +
\mathrm{near} (c \xi)$, where $c$ is a parameter that allows varying the
width of the delay distribution; $\xi$ is Gaussian distributed with zero
mean and standard deviation one; $\mathrm{near}$ denotes the nearest
integer. Depending on $\tau_0$ and $c$ the distribution of delays has to
be truncated to avoid negative delays.  Since the focus of this paper is
the influence of the array topology and the connectivity we keep the
random delays Gaussian distributed with $\langle \tau_{ij} \rangle
 \sim \tau_0=5$ and $c=2$. The numerical results are summarized in figures
\ref{fig-cpr}-\ref{fig-strong}.

In Fig. \ref{fig-cpr} we can see density plots of $\sigma^2$ as a
function of the mean number of links per node, $b =1/N\sum_{i=1}^N b_i$
and $\epsilon$. The four different panels correspond to the different
networks mentioned above. Despite the differences for small number of
neighbors, we observe that the synchronizability is largely independent
of the topology.  Furthermore, the similarity between
Figs.~\ref{fig-cpr}(c) and (d) clearly suggest that the
synchronizability does not depend on the average path length (defined as
the distance between two nodes, averaged over all pairs of nodes). 

For weak coupling (roughly speaking, $\epsilon \lesssim 0.4$), the array
does not synchronize regardless the number of neighbors and
topology. However, there are worth noting different behaviors depending
on the value of $\epsilon$. To gain additional insight, it is shown in
Fig.~\ref{fig-weak} the value of $\sigma^2$ as a function of b for
different values of $\epsilon$.  We observe that for $\epsilon=0.02$ (a)
the value of $\sigma^2$ decreases monotonously as $b$ increases, to a
limiting non-zero value. For $\epsilon=0.04$ (b) there is still no
synchronization for any connectivity value, but there is a
non-monotonous dependence of $\sigma^2$ with $b$, that reveals the
existence of an optimal number of neighbors for which there is an
enhancement of the array propensity for synchronization. Finally, in (c)
we observe that, for $\epsilon=0.6$, in sharp contrast to (a),
$\sigma^2$ increases monotonously before reaching a non-zero limiting
value.

On the other hand, for strong enough coupling ($\epsilon \gtrsim 0.4$) the
array synchronizes in the homogeneous steady state if the number of the
neighbors is large enough.  In figure \ref{fig-strong} we observe
$\sigma^2$ as functions of $b$ for different values of $\epsilon$.
Depending on the value of $\epsilon$ there is a minimum value of
neighbors required to synchronize the array. Moreover, this critical
value decreases with increasing $\epsilon$.

In summary, we studied the synchronization of coupled maps in complex
networks with time-delayed interactions focusing on the influence of
array connectivity and topology. For weak coupling no synchronization
was found, but an enhancement of the synchronization propensity was
observed for particular connectivity values. On the other hand, for
strong coupling there is synchronization provided that the number of
neighbors was large enough.

\end{document}